\documentstyle[eqsecnum,aps,prd]{revtex}
\def\and{{$\&$ }}
\def\lsim{\mathrel{\mathpalette\oversim<}}
\def\gsim{\mathrel{\mathpalette\oversim>}}
\def\oversim#1#2{\lower0.2ex\vbox{\baselineskip0pt\lineskip0pt
  \lineskiplimit0pt\ialign{$#1\hfil##\hfil$\crcr#2\crcr\sim\crcr}}}
\begin{document}
\draft


\parbox{\hsize}{
\begin{flushright}
HUPD-9813, KUNS-1507, YITP-98-42
\\
May 1998
\\
Revised July 1998
\end{flushright}

\title{Small-Scale Fluctuations in Cosmic X-ray Background : \\
A Power Spectrum Approach}
\author{Kazuhiro Yamamoto}
\address{Department of Physics, Hiroshima University,
 Higashi-Hiroshima 739-8526, Japan\\
Yukawa Institute for Theoretical Physics, Kyoto University,
 Kyoto 606-8502, Japan}

\author{Naoshi Sugiyama}
\address{Department of Physics, Kyoto University, Kyoto,
  606-8502, Japan}
\maketitle
}


\vspace{10pt}

\begin{abstract}
Equations to investigate fluctuations in cosmic X-ray background 
radiation due to point-like sources at high-redshift are 
formulated in a systematic way.
The angular power spectrum of X-ray background fluctuations
is investigated from large-scales to small-scales in various 
cosmological models such as open universe models and models
with the cosmological constant, 
assuming a simple evolution model of the sources.
The effect of epoch-dependent bias is demonstrated for small-angle
fluctuations. The contribution from shot noise fluctuations
is also discussed.  

 
\end{abstract}

\thispagestyle{empty}
\def\bfx{{\bf x}}
\def\nuz{{\nu_0}}
\def\calH{{\cal H}}
\def\calL{{\cal L}}
\def\Inu{{I_\nu}}
\def\fnu{{f}}
\def\I{{I}}
\def\barm{{m}}
\def\tildek{{\tilde k}}
\def\tildeo{{\omega}}
\def\etac{{\eta_c}}
\section{Introduction}
Cosmic X-ray background (CXB) has been studied for long time in
astrophysics\cite{Fabian}.  Recently several authors have discussed the
possibility of the CXB fluctuations as a probe of the structure
formation in the high redshift
universe\cite{Lahav,Treyer,Boughn,Carrera,Barcons,TL,WLR}.  These works are
motivated from observational results of the deep survey carried out by
ROSAT, which resolved a significant fraction of the X-ray background
into sources ($\sim 70\%$). It turns out to be that most of the sources are
extra-galactic objects, i.e., AGN and other X-ray luminous galaxies in
the high redshift universe.  While situation is not so clear in the harder
band, something similar would be happening.  Several X-ray mission
projects are in progress\cite{mission}, which will give us a fine solution
of the source problem in near future.

On the other hand, a scenario of the cosmic structure
formation is now constrained from the observations, e.g.,
temperature anisotropies in the cosmic microwave background
radiation (CMB) and large-scale distribution of galaxies. 
Future satellite experiments of CMB and survey projects
of the large-scale distribution of galaxies will provide severer 
constraints on theoretical models of the large scale structure formation.
At present the scenario introducing Cold Dark Matter (CDM) seems 
to be successful though some modification may be required\cite{WSSD,WS}.

Formation processes of subgalactic objects, galaxies, AGNs and clusters
at the high redshift universe are one of the hottest topics in the area
of astrophysics and cosmology.  Since these objects are parts of 
the large scale structure of the universe, 
we could expect that information about the
structure formation in the high redshift universe would be obtained 
by investigating  them.


Several  works have been done to answer following questions:
Which kind of information about cosmic structure formation can be obtained 
from an analysis of the X-ray background fluctuations? 
Are they really a probe of cosmology?   For example, 
Boughn, et al. claimed that the cross correlation between the X-ray 
background fluctuations and the microwave background anisotropies 
can provide a constraint on a cosmological constant.
Small-scale fluctuations in the X-ray 
background have been investigated in \cite{Carrera,Barcons,MM,DeZotti,BF}.
On the other hand, Lahav et al. \cite{Lahav} 
calculated the angular power spectrum of 
the fluctuations by a multipole expansion method.
They mainly focused on 
large scale fluctuations and their formula is only limited to the case 
of the flat universe. 
Recently Treyer, et al. have compared the theoretical models
by Lahav et al. with the observational data of HEAO1\cite{Treyer}.

In this paper we expand Lahav et al.'s formula to more general
cosmological models, i.e., open universe models and models with the
cosmological constant, and investigate general behaviors of the angular
power spectrum of fluctuations from large-scales to small-scales.  
The X-ray background fluctuation in the open universe has been considered
in ref.\cite{Miura}, and the similar expression was obtained.
However, in that paper, the discussion was only focused on 
the large-scale fluctuations due to the source clustering,
and the derivation of equations was very complicated. 
In this paper we formulate the treatment in a simple way.
And the shot noise fluctuation is treated in a systematic way.
The formulation developed in this paper will be useful for studying the
clustering of high redshift sources not only for X-ray sources but also
for other point-like sources such as radio galaxies \cite{Loan,Vall} and 
$\gamma$-ray bursts.  

This paper is organized as follows.  In section 2 we derive expression
for the angular two-point correlation function by using the multipole
expansion method in a systematic way.  In section 3 the fluctuations in
the X-ray background are investigated assuming a simple luminosity
function of X-ray sources. The competition of the fluctuations due to
the source clustering and the shot noise is also studied there.
Section 4 is devoted to summary and discussions.  Throughout this paper
we use the unit $c=1$.

\def\log{{\rm log}}
\def\Sc{{S_c}}
\def\zmax{z_{\rm max}}
\def\zmin{z_{\rm min}}
\section{Formulation}
In this section we formulate equations to calculate background 
fluctuations which come from point-like sources at high redshift.
Essence of the formulation is divided into two parts.
One is how the sources are distributed in the universe.
This is the problem of source evolution and statistics of 
the distribution, which we describe in the latter part
of this section. 
The other part is how the sources are observed in the sky.
Which is obtained by solving light propagation in the expanding 
universe, if the distribution of the sources in the universe was
given.

First we consider the light propagation.
The photon intensity $\Inu$ in an expanding universe
is described by the radiative transfer equation:
\begin{equation}
  {\partial\Inu \over\partial\eta}
  +\gamma^j{\partial\Inu \over\partial x^j}
  +\calH\biggl[3\Inu-\nu{\partial\Inu \over\partial \nu}\biggr]
  +{d\gamma^j \over d\eta}{\partial\Inu \over\partial \gamma^j}
  =a j_\nu(\bfx,\eta), 
\label{RT}
\end{equation}
where $a$ is the scale factor which is normalized 
to be unity at present,  
$\eta$ is conformal time defined by $a d\eta=dt$ with the cosmic time
$t$, 
$\calH$ is the Hubble parameter defined as ${\dot a}/ a$ with 
over-dot denoting $\eta$ differentiation,
$\gamma^i$ is the directional vector of the photon momentum, 
and  $j_\nu(\bfx)$ is a field of emissivity
per unit comoving volume. 
Assuming that an ($i$)-th point-like source is 
located at the coordinate $\bfx^{(i)}$ and has the 
power put out per unit frequency and per unit time
$L_\nu^{(i)}(\eta)$,
we write the field  of emissivity as
\begin{equation}
  j_\nu(\bfx,\eta)={1\over 4\pi a^3}\sum_{i} L_\nu^{(i)}(\eta)
  \delta^{(3)}(\bfx-\bfx^{(i)}).
\label{emissivity}
\end{equation}

We solve equation (\ref{RT}) with (\ref{emissivity}) in 
the Friedmann-Robertson-Walker space-time with the line element
\begin{equation}
  ds^2=-dt^2+a^2\gamma_{ij}dx^idx^j ~ ,
\label{metric}
\end{equation}
where $\gamma_{ij}$ is the three-metric on a space of 
constant negative curvature $K$:
\begin{equation}
  \gamma_{ij}dx^idx^j={1\over -K}\Bigl(d\chi^2+\sinh^2\chi(d\theta^2+\sin ^2\theta d\phi^2)\Bigr).
\label{hyperelement}
\end{equation}
In order to take the limit of the flat universe,
let us first introduce a radial coordinate $r$ instead of $\chi$
by $\chi=\sqrt{-K} r$ and take the limit $K\rightarrow 0$.
The scale factor $a$ is governed by the Friedmann equation,
\begin{equation}
  \calH^2=\biggl({\dot a\over a}\biggr)^2=H^2_0
  \biggl( {1\over a}\Omega_0+\Omega_K+{a^2} \Omega_\Lambda\biggr)~,
\label{FriedmannEq}
\end{equation}
where $H_0$ is the Hubble parameter with $H_0=100~h~{\rm km/s/Mpc}$,
$\Omega_0$ is the density parameter, $\Omega_\Lambda \equiv
\Lambda/3H_0^2$ 
which is the density parameter of the cosmological constant $\Lambda$, 
and $\Omega_K \equiv 1-\Omega_0-\Omega_\Lambda$ which describes the spatial 
curvature of the universe.

Employing a  new variable $\fnu=\Inu/\nu^3$, we can rewrite
equation (\ref{RT}) as 
\begin{equation}
  {\partial\fnu \over\partial\eta}
  +\gamma^j{\partial\fnu \over\partial x^j}
  -\calH\nu{\partial\fnu \over\partial \nu}
  +{d\gamma^j \over d\eta}{\partial\fnu \over\partial \gamma^j}
  ={a\over4\pi (a\nu)^3}\sum_i L^{(i)}_\nu(\eta) 
  \delta^{(3)}(\bfx-\bfx^{(i)}).
\label{RTB}
\end{equation}
Introducing  $\nuz(=\nu a)$ instead of $\nu$, (\ref{RTB}) reduces to
\begin{equation}
  {\partial\fnu \over\partial\eta}
  +\gamma^j{\partial\fnu \over\partial x^j}
  +{d\gamma^j \over d\eta}{\partial\fnu \over\partial \gamma^j}
  ={a\over 4\pi \nuz^3}\sum_i 
  L^{(i)}_{\nu\rightarrow \nuz/a}(\eta) \delta^{(3)}(\bfx-\bfx^{(i)}). 
\label{REC}
\end{equation}
Operating $\int_{\nu_1}^{\nu_2}d\nuz \nuz^3$ on the both side of 
equation (\ref{REC}), we have 
\begin{equation}
  {\partial\I \over\partial\eta}
  +\gamma^j{\partial\I \over\partial x^j}
  +{d\gamma^j \over d\eta}{\partial\I \over\partial \gamma^j}
  ={a\over 4\pi }\sum_i \delta^{(3)}(\bfx-\bfx^{(i)})
  \int_{\nu_1}^{\nu_2} d\nuz L^{(i)}_{\nu\rightarrow \nuz/a}(\eta), 
\label{RED}
\end{equation}
where $I=\int_{\nu_1}^{\nu_2}d\nuz \nuz^3 f$.
Since equation (\ref{RED}) is rewritten as
\begin{equation}
  {d I(\eta,\bfx,\vec\gamma)\over d\eta}
  ={\partial\I \over\partial\eta}
  +\gamma^j{\partial\I \over\partial x^j}
  +{d\gamma^j \over d\eta}{\partial\I \over\partial \gamma^j}
  ={a\over 4\pi }\sum_i \delta^{(3)}(\bfx-\bfx^{(i)})
  K^{(i)}(\eta) L^{(i)}(\eta),
\label{REE}
\end{equation}
where the luminosity of the $i$-th source is defined by
\begin{equation}
  L^{(i)}(\eta)=\int_{\nu_1}^{\nu_2}d\nu L_\nu^{(i)} (\eta),
\label{L}
\end{equation}
and $K^{(i)}(\eta)$ is defined by 
\footnote{$K^{i}(\eta)/a(\eta)$ denotes the usual K-correction.}
\begin{eqnarray}
  &&
  K^{(i)}(\eta) ={1\over L^{(i)}(\eta)} 
  \int_{\nu_1}^{\nu_2} d\nuz L^{(i)}_{\nu\rightarrow \nuz/a}(\eta),   
\label{K}
\end{eqnarray}
then by integrating (\ref{REE}), we get
\begin{equation}
  I(\eta_0,\bfx_0,\vec\gamma)= {1\over 4\pi} \sum_i \int d\eta a(\eta)  
  K^{(i)}(\eta) L^{(i)}(\eta)
  \delta^{(3)}(\bfx(\eta,\vec\gamma)-\bfx^{(i)}),
\label{REFB}
\end{equation}
where $\bfx(\eta,\vec\gamma)$ stands for a photon path.

Because $\sum_i\delta^{(3)}(\bfx-\bfx^{(i)})$ is regarded as a number
density field and we replace it with
\begin{equation}
  \sum_i\delta^{(3)}(\bfx-\bfx^{(i)}) \Longrightarrow 
  \int d\log L n(L,\eta,\bfx),
\end{equation}
where $n(L,\eta,\bfx)$ is the luminosity function which 
denotes a number of sources per unit comoving volume
and per unit $\log L$. Then we rewrite (\ref{REFB}) as
\begin{equation}
  I(\eta_0,\bfx_0,\vec\gamma)= 
  {1\over 4\pi} \int d\log L L  \int d\eta a(\eta) K(L,\eta) 
   n(L,\eta,\bfx(\eta,\vec\gamma)),
\label{REFC}
\end{equation}
where $K(L,\eta)$ is defined for a source with luminosity 
$L$ in same way as (\ref{K}).

Here we mention the range of $\eta$ integration in equation
($\ref{REFC}$). The range of $\eta$ integration depends on observational
situation and strategy.  In this paper we focus on background
fluctuations.  In order to reduce the shot noise fluctuations, which are
described later in this paper, we assume that bright nearby sources are
removed from an observed map.  Darker and darker sources we get rid of,
more and more distant sources are removed.  This flux cutoff limit is
denoted by $\Sc$.  The local luminosity of a source $L$ is related with
the observed flux $S$ and the distance, which we describe by the
conformal time $\eta$ in stead of the redshift, as
\begin{equation}
  S= {a L K(L,\eta)\over 4\pi D(\eta)^2},  
\label{Scut}  
\end{equation}
where $D(\eta)=(-K)^{-1/2} \sinh(\sqrt{-K}(\eta_0-\eta))$ and $\eta_0$ is
the conformal time at the present epoch. 
Note that the distance $D$ is related with the 
luminosity distance $d_L$ by $D(\eta)=d_L/(1+z)$ \cite{SKL}.
In case of observations with the flux cutoff limit $\Sc$, the range of 
$\eta$ integration must be $0\leq\eta\leq\eta_{\rm c}$, 
where $\eta_{\rm c}$ is determined by solving (\ref{Scut}) 
with setting $S=\Sc$. Thus $\eta_{\rm c}$ is a function of 
$L$ and $\Sc$, in general.
As we will see in the next section, the removal of bright sources 
is an important factor in predicting the fluctuations.

In order to discuss statistics of the fluctuations
we define the angular two-point correlation function:
\begin{eqnarray}
  C(\theta)&=&
\int\int 
  {d\Omega_{\vec\gamma_1}\over 4\pi}
  {d\Omega_{\vec\gamma_2}\over 2\pi}
  \delta(\vec\gamma_1\cdot\vec\gamma_2-\cos\theta)
  I(\eta_0,\bfx_0,\vec\gamma_1)I(\eta_0,\bfx_0,\vec\gamma_2)
\nonumber
\\
  &=&\int\int 
  {d\Omega_{\vec\gamma_1}\over 4\pi}
  {d\Omega_{\vec\gamma_2}\over 2\pi}
  \delta(\vec\gamma_1\cdot\vec\gamma_2-\cos\theta)
\nonumber
\\
  &&\times 
   {1\over 4\pi}\int d\log L_1 L_1 \int^{\etac}_0 d\eta_1 {a_1} K(L_1,\eta_1) 
  ~n(L_1,\eta_1,\bfx(\eta_1,\vec\gamma_1))
\nonumber
\\
  &&\times 
   {1\over 4\pi}\int d\log L_2 L_2 \int^{\etac}_0 d\eta_2 {a_2} K(L_2,\eta_2) 
  ~n(L_2,\eta_2,\bfx(\eta_2,\vec\gamma_2))
\label{CA}
\end{eqnarray}
where $a_1=a(\eta_1)$ and $a_2=a(\eta_2)$.

Theoretically we can only predict the ensemble average of fluctuations.
To obtain the ensemble average of the
two-point correlation function, we need the ensemble
average $\bigl<n(L_1,\eta_1,\bfx_1) n(L_2,\eta_2,\bfx_2)\bigr>$
from equation (\ref{CA}). 
In case of point-like sources, it is well known that
the ensemble average of a product of the density field   
has three terms, i.e., homogeneous term, clustering term and shot term
\cite{Bertschinger,Peebles}.
As we are dealing with the density field as a function of
luminosity and time, the relation 
$\bigl<n(L_1,\eta_1,\bfx_1) n(L_2,\eta_2,\bfx_2)\bigr>$
is not trivial in a strictly sense.
However, assuming that the luminosity of sources is statistically 
independent of their positions relative to the other sources,
we set the following relation 
\cite{MM,DeZotti}: 
\begin{eqnarray}
  &&\bigl<n(L_1,\eta_1,\bfx_1) n(L_2,\eta_2,\bfx_2)\bigr>=
  \bar n(L_1,\eta_1)\bar n(L_2,\eta_2)
  +\bar n(L_1,\eta_1)\bar n(L_2,\eta_2)\xi(\eta_1,\eta_2;\bfx_1-\bfx_2)
\nonumber
\\
  &&\hspace{1.0cm}
   + \bar n(L_1,\eta_1)\delta(\log L_1-\log L_2)\delta^{(3)}(\bfx_1-\bfx_2),
\label{NN}
\end{eqnarray}
where $\bar n(L,\eta)$ is the spatially averaged quantity of 
$n(L,\eta,\bfx)$.
The first term of r.h.s. of (\ref{NN}) is the isotropic background 
component, the second term describes the clustering 
of the sources, the third term is called the shot term, which arises
from the discreteness of the sources \cite{Bertschinger}.
Then the ensemble average $\bigl<C(\theta)\bigr>$ has 
three terms, i.e., isotropic background term, clustering term 
and shot noise term, which we write as
$\bigl<C(\theta)\bigr>=\bigl<C^{\rm ISO}\bigr>+\bigl<C^{\rm CL}(\theta)\bigr>
+\bigl<C^{\rm SN}(\theta)\bigr>$.

\subsection{Isotropic background}
First we consider the isotropic background term. From (\ref{CA}) 
and the first term of r.h.s. of (\ref{NN}), 
we easily get 
\begin{equation}
  \bigl<C^{\rm ISO}\bigr>=
  \biggl({1\over 4\pi}\int d\log L L \int^{\etac}_0 d\eta a K(L,\eta) 
  \bar n(L,\eta)\biggr)^2 =\Bigl(I^{(0)}\Bigr)^2,
\end{equation}
where we used 
\begin{equation}
  \int\int
  {d\Omega_{\vec\gamma_1}\over 4\pi}
  {d\Omega_{\vec\gamma_2}\over 2\pi}
  \delta(\vec\gamma_1\cdot\vec\gamma_2-\cos\theta)=1.  
\end{equation}

\subsection{Fluctuations from Source Clustering}
The correlation function of the source clustering term is
\begin{eqnarray}
 &&\bigl<C(\theta)^{\rm CL}\bigr>
  =\int\int 
  {d\Omega_{\vec\gamma_1}\over 4\pi}
  {d\Omega_{\vec\gamma_2}\over 2\pi}
  \delta(\vec\gamma_1\cdot\vec\gamma_2-\cos\theta)
\nonumber
\\
 &&{\hspace{1cm}}\times 
   {1\over 4\pi}\int d\log L_1 L_1 \int^{\etac}_0 d\eta_1 {a_1} K(L_1,\eta_1) 
  ~\bar n(L_1,\eta_1)
\nonumber
\\
  &&{\hspace{1cm}}\times 
   {1\over 4\pi}\int d\log L_2 L_2\int^{\etac}_0 d\eta_2 {a_2} K(L_2,\eta_2) 
  ~\bar n(L_2,\eta_2)
\nonumber
\\
  &&{\hspace{1cm}}\times 
  ~\xi(\eta_1,\eta_2;\bfx_1(\eta_1,\vec\gamma_1)-\bfx_2(\eta_2,\vec\gamma_2)).
\label{CB}
\end{eqnarray}
To relate the correlation function of X-ray sources with
that of CDM, we write
\begin{equation}
  \xi(\eta_1,\eta_2;\bfx_1-\bfx_2)
  =b_X(\eta_1)b_X(\eta_2)\xi(\eta_1,\eta_2;\bfx_1-\bfx_2)_{\rm CDM},
\end{equation}
where $b_X(\eta)$ is the bias factor\footnote{
The bias factor could be a function of luminosity $L$ in general.
However we ignore the dependence of luminosity 
for simplicity.},
the CDM density correlation function is 
$\xi(\eta_1,\eta_2;\bfx_1-\bfx_2)_{\rm CDM}
=\bigl<\delta_{\rm c}(\eta_1,\bfx_1)\delta_{\rm c}(\eta_2,\bfx_2)\bigr>$,
and $\delta_{\rm c}(\eta,\bfx)$ is the CDM density contrast.
While the nonlinearity of the CDM density contrast may be
effective, however, we only consider linear theory in this paper.
In this case $\xi(\eta_1,\eta_2;\bfx_1-\bfx_2)_{\rm CDM}$ is
proportional to $D_1(\eta_1)D_1(\eta_2)$, where $D_1(\eta)$ is
the linear growth rate normalized to unity at present.

We next consider to express the CDM density contrast in terms of
the density power spectrum. In order to express Gaussian random process
of the CDM density field, we introduce probability variables
$a_{\tildeo lm}$, which satisfy,
\begin{equation}
  \bigl<a_{\tildeo lm}\bigr>=0,
  \hspace{1cm}
  \bigl<a_{\tildeo lm}a^{*}_{\tildeo'l'm'}\bigr>
  =\delta(\tildeo-\tildeo')\delta_{ll'}\delta_{mm'}.
\end{equation}
For the condition of reality of density field 
we assume $a^{*}_{\tildeo lm}=a_{\tildeo l-m}$.
Then we can write the CDM density fluctuation fields as
\begin{equation}
  \delta_{\rm c}(\eta,\bfx)=\int_0^\infty d\tildeo \sum_{lm} 
  a_{\tildeo lm} \delta_{\tildeo}(\eta){\cal Y}_{\tildeo lm}(\bfx),
\end{equation}
where $\delta_\tildeo(\eta)$ is the root-mean-square of the 
coefficient associated with the orthonormalized harmonics 
in the hyperbolic universe 
\begin{eqnarray}
    {\cal Y}_{\tildeo lm}(\bfx)
  &=& (-K)^{3/4} W_{\tildeo l}(\chi) Y_{lm}(\Omega)
\nonumber
  \\
  &=& (-K)^{3/4} \biggl\vert
  {\Gamma(i\tildeo+l+1)\over\Gamma(i\tildeo)}\biggr\vert
  {P^{-l-1/2}_{i\tildeo-1/2}(\cosh\chi)\over \sqrt{\sinh\chi}}
  Y_{lm}(\Omega),
\label{Wlm}
\end{eqnarray}
which satisfies
\begin{equation}
  \int (-K)^{-3/2} \sinh^2\chi d\chi d\Omega
  {\cal Y}_{\tildeo lm}(\chi,\Omega) 
  {\cal Y}^{*}_{\tildeo' l'm'}(\chi,\Omega)
  =\delta(\tildeo-\tildeo')
   \delta_{ll'}\delta_{mm'},
\end{equation}
where $\Gamma(z)$ is the Gamma function and $P^\nu_\mu(z)$ is the
Legendre function. Here $\tildeo$ is a non-dimensional wave number 
and ${K}(\tildeo^2+1)$ is the eigen-value of the scalar 
harmonics ${\cal Y}_{\tildeo lm}(\bfx)$ in the
hyperbolic universe (\ref{hyperelement}) \cite{Gouda,HSb,Wilson}.
Then we have 
\begin{equation}
  \bigl<\delta_{\rm c}(\eta_1,\bfx_1)\delta_{\rm c}(\eta_2,\bfx_2)\bigr>
  =\sum_{lm}\int_0^\infty d\tildeo 
  \delta_\tildeo(\eta_1){\cal Y}_{\tildeo lm}(\chi_1,\Omega_1)
  \delta_\tildeo(\eta_2){\cal Y}^{*}_{\tildeo lm}(\chi_2,\Omega_2).
\end{equation}

     From a straightforward calculation using 
the relations
\begin{eqnarray}
  &&\delta(\vec\gamma_1\cdot\vec\gamma_2-\cos\theta)
  =\sum_{l=0}{2l+1\over 2}P_l(\vec\gamma_1\cdot\vec\gamma_2) P_l(\cos\theta),
\label{relationA}
\\
  &&P_l(\vec\gamma_1\cdot\vec\gamma_2)={4\pi\over 2l+1}
  \sum_{\barm=-l}^{\barm=l}
  Y_{l\barm}(\Omega_{\vec\gamma_1})Y^{*}_{l\barm}(\Omega_{\vec\gamma_2}),
\label{relationB}
\end{eqnarray}
where  $Y_{lm}(\Omega_{\vec\gamma})$ is the spherical harmonics 
in an unit sphere, 
equation (\ref{CB}) reduces to
\footnote{
Throughout this paper we consider an ideal detector
with infinite small angular resolution. The effect of
a finite beam width and a limited observational area can
be taken into account by multiplying a window function 
in the similar way as the case of CMB temperature anisotropies.
}
\begin{equation}
  \bigl<C(\theta)^{\rm CL}\bigr>=\sum_l{2l+1\over 4\pi} C^{\rm CL}_l P_l(\cos\theta),
\label{Cl}
\end{equation}
with
\begin{eqnarray}
  C^{\rm CL}_l=\int_0^\infty d\tildeo (-K)^{3/2} 
  \biggl({1\over 4\pi} \int d\log L L 
  \int^{\etac}_0 d\eta a K(L,\eta)\ \bar n(L,\eta)
  b_X(\eta)\delta_\tildeo(\eta) W_{\tildeo l}(\Delta\eta)
  \biggr)^2,
\end{eqnarray}
where $\Delta\eta=\sqrt{-K}(\eta_0-\eta)$ and
$W_{\tildeo l}(\chi)$ is defined in equation (\ref{Wlm}).
This expression is rewritten as \cite{Miura}
\begin{eqnarray}
  C^{\rm CL}_l={2\over \pi}\int_0^\infty d\tildek \tildek^2
  M_l 
  \biggl({1\over 4\pi}\int d\log L L 
  \int^{\etac}_0 d\eta a K(L,\eta) \bar n(L,\eta)
  b_X(\eta)\delta_\tildeo(\eta) X_{\tildeo}^l(\Delta\eta)
  \biggr)^2,
\label{CCLA}
\end{eqnarray}
where $M_l=(\tildek^2-K)\cdots (\tildek^2-l^2K)/(\tildek^2-K)^l$, 
$\tildek=\sqrt{-K}\tildeo$, and 
\begin{equation}
  X_\omega^{l}(\chi)=\biggl({\pi(\tildeo^2+1)^l \over 2}\biggr)^{1/2}
  {P^{-l-1/2}_{i\omega-1/2}(\cosh\chi)\over \sqrt{\sinh\chi}}~.
\label{CCLB}
\end{equation}
In the limit of flat universe, i.e., $K\rightarrow0$, 
the function $X_\omega^{l}(\chi)$ reduces to the spherical 
Bessel function \cite{Gouda,HSb,Wilson}.

\subsection{Random Fluctuations}
Finally we consider random fluctuations from the shot term.
The ensemble average of the two-point correlation function is 
\begin{eqnarray}
  \bigl<C(\theta)^{\rm SN}\bigr>
  &=&\int\int 
  {d\Omega_{\vec\gamma_1}\over 4\pi}
  {d\Omega_{\vec\gamma_2}\over 2\pi}
  \delta(\vec\gamma_1\cdot\vec\gamma_2-\cos\theta)
\nonumber
\\
  &&\times 
   {1\over 4\pi}\int d\log L_1 L_1 \int^{\etac}_0 d\eta_1 {a_1} K(L_1,\eta_1) 
   {1\over 4\pi}\int d\log L_2 L_2 \int^{\etac}_0 d\eta_2 {a_2} K(L_2,\eta_2) 
\nonumber
\\
  &&\times 
  ~\bar n(L_1,\eta_1) \delta(L_1-L_2)
  ~\delta^{(3)}(\bfx_1(\eta_1,\vec\gamma_1)-\bfx_2(\eta_2,\vec\gamma_2)).
\label{ClShot}
\end{eqnarray}
As the delta function is expressed
\begin{equation}
  \delta^{(3)}(\bfx_1-\bfx_2)= (-K)^{3/2} 
  {\delta(\chi_1-\chi_2)\over \sinh^2\chi_1}
  \delta^{(2)}(\vec\gamma_1-\vec\gamma_2),
\end{equation}
then equation (\ref{ClShot}) leads to
\begin{equation}
  C^{\rm SN}(\theta)=\sum_l{2l+1\over 4\pi} C^{\rm SN}_l P_l(\cos\theta),
\end{equation}
with
\begin{eqnarray}
  C^{\rm SN}_l=\int d\log L L^2 \int^{\etac}_0 d\eta 
  \biggl({1\over 4\pi}a K(L,\eta)\biggr)^2 \bar n(L,a)
  {-K\over \sinh^2 \Delta\eta},
\label{SNfluct}
\end{eqnarray}
where $\Delta\eta=\sqrt{-K}(\eta_0-\eta)$.  
Here we used equations (\ref{relationA}) and (\ref{relationB}).
This spectrum is independent of $l$, which
implies that this fluctuation is a white noise type.
And this means $ \bigl<C(\theta)^{\rm SN}\bigr>\propto \delta(\theta)$.

\section{Fluctuations in a simple model}
In this section we discuss characteristic behavior of
fluctuations by considering a simple evolution
model of X-ray sources. 
The evolution and population of X-ray sources at high redshift 
are less well understood.
Here we consider a model of a single class of sources,
i.e., the case that the sources have same luminosity profile
at a cosmic time.
In this case the luminosity function is given by
\begin{equation}
  \bar n(L,\eta)= \bar n(a) \delta(\log L-\log {\cal L}),
\end{equation}
where $\cal L$ is the luminosity of a source
and $\bar n$ is the averaged number density of sources.
We assume that the sources are distributed to the redshift 
$\zmax$. 
Furthermore we assume power-law evolution 
of the source number density and the luminosity:
\begin{equation}
  \bar n(a)=n_0 a^{-d}, 
  \hspace{1cm}
  {\cal L} = L_0 a^{-e},
\end{equation}
where $n_0$ and $L_0$ are constants. We further assume 
the power-law frequency dependence $L_\nu\propto \nu^{-\alpha}$,
then we have
\begin{equation}
  K(L,\eta)= a(\eta)^{\alpha}.
\end{equation}
As for the bias, we take into account 
the epoch-dependent bias
by setting \cite{Treyer,Fry}
\begin{equation}
  b_X(\eta)=b_{X0}+z[b_{X0}-1],
\label{bX}
\end{equation}
where $z$ is the redshift and $b_{X0}$ is a constant. 
Although this simple formula might be 
applicable  only in the Einstein de Sitter universe,
we use it even for other cosmological models for simplicity.

\subsection{Fluctuations from Source Clustering}
The above simplification leads to  the following simple expressions
for the isotropic background and the angular power spectrum
of source clustering:
\begin{eqnarray}
  &&I^{(0)}={n_0L_0\over 4\pi} \int^{\etac}_{\eta_{i}}d\eta a^{1-d-e+\alpha},
\\
  &&C_l^{\rm CL}={2\over \pi}\int_0^\infty d\tildek\tildek^2 M_l
  \biggl({n_0L_0\over 4\pi}\int^{\etac}_{\eta_{i}}d\eta
  a^{1-d-e+\alpha} 
  b_X(\eta)
  \delta_\tildeo(\eta) X_\tildeo^l(\Delta\eta) \biggr)^2,
\end{eqnarray}
where $\eta_i$ is the conformal time at the redshift 
$\zmax$. Note also that the relative amplitude of fluctuations 
$C_l^{\rm CL}/(I^{(0)})^2$ is independent of $n_0L_0$.  
The evolution of the X-ray sources are less understood. 
Let us consider the case that the volume emissivity
evolves with the time, and we set $p\equiv d+e=3$
for simplicity\cite{Lahav}.
We adopt this value and $\alpha=0.4$ in the rest of this paper \cite{Lahav}.

The angular power spectrum is shown in Figure~1 for various 
cosmological models. In Fig.1(a) and 1(b), the three panels 
correspond to the different cosmological models, a 'standard' 
CDM model (SCDM), a CDM model with the  cosmological 
constant ($\Lambda$CDM) and an open CDM model (OCDM) \cite{YB}. 
In each panel three
lines stand for the different parameters for removal of bright sources.
For simplicity we take this effect into account by specifying
the distance of redshift $\zmin$ instead of specifying $\Sc$. 
We assumed that the sources of $0\leq z\leq\zmin$ were removed.
In the figure three lines correspond to the cases $\zmin=0,~0.02,~0.1$.
Fig.1(a) is the case of no bias by setting $b_{X0}=1$ in 
equation (\ref{bX}). In Fig.1(b) we chose $b_{X0}=1.6$ as
a case of the epoch-dependent bias, where the bias factor becomes 
significant at $z\gsim1$.
The CDM density power spectrum is normalized as $\sigma_8=1$. 

First notable feature is that the amplitude of fluctuations
strongly depends on $\zmin$ for low multipoles, i.e., large-angle
scales. As $\zmin$ controls distance to which nearby 
sources are removed, the decrease of the low multipoles 
is due to the removal of the sources.
This implies that the nearby sources dominantly 
contribute to the low multipoles \cite{Miura}.
On the contrary, the high multipoles 
do not depend on $\zmin$. 

In order to understand dominant contributors of sources to fluctuations
in more detail, we show the amplitude of the multipoles
$l=1,~10,~30,~100$, as a function of $\zmin$ in Fig.2.  The large value
of $\zmin$ taken in this figure might be unrealistic from the
observational view point for an all sky survey. However this figure is
instructive to understand what sources are dominant contributors to the
fluctuations of each angular size.  The steep decrease of the low
multipole at $\zmin\lsim 0.1$ indicates that low redshift sources
sensitively contribute to the large-angle fluctuations.  On the other
hand the higher multipole ($l=100$) does not sensitively depend on
$\zmin$ and it is not almost affected by the removal of the sources for
$\zmin\lsim2$. This means that the high redshift sources of $z\gsim1$
are the dominant contributors to the small-scale fluctuations
$l\gsim100$.  It is shown that the amplitude of the fluctuations
increases for $\zmin\gsim2$. The width of the range in which sources are
distributed becomes thin when $\zmin$ becomes near $\zmax$. This effect
makes the amplitude of the fluctuations increase since the fluctuations 
suffer less significant thickness damping, that is contributions  from 
sources at different redshifts cause the cancellation.   We think that the
increase of the fluctuations at $\zmin\gsim2$ in Fig.2 are due to this
effect.

Comparing Fig.1(a) and 1(b), evolution of 
the bias derives significant difference on small scales.
For the low multipoles, since the nearby sources of low redshift 
dominantly contribute to the fluctuations, 
the (low multipole) amplitude scales by the bias factor
at present epoch $b_{X0}$. 
However the amplitude on small scales is 
significantly increased for the epoch-dependent bias model.
This is understood that the high redshift sources 
are the dominant contributors to the small-angle fluctuations
and the amplitude is increased due to the large value 
of the bias factor at high redshift.

It is notable that the dependence on the cosmological models is quite 
weak even on small scales. Because we have employed $\sigma_8$ normalization,
the shape of the density power spectrum of the OCDM model is almost same
as the $\Lambda$CDM model. The difference between these two models is
small at a few$\times10\%$ level.  However the amplitude of fluctuations
of the OCDM model is slightly larger than the $\Lambda$ model on small
scales.  This would be understood as follows.  When we see a same
physical size in the open universe, the angle size becomes smaller
compared with the case of the flat universe due to the curvature of
background geometry.  Therefore the large-scale fluctuations in the open model
shift to the smaller scale. This is the same situation as the case of CMB
anisotropies.

\subsection{Shot Noise v.s. Source Clustering }

In this subsection we discuss the shot noise fluctuations
comparing them with the fluctuations due to the source clustering. 
The shot noise fluctuations are random fluctuations
originated from discreteness of the sources.
The spectrum is the white noise type and the 
angular power spectrum does not depend on $l$.
In our simple model, equation (\ref{SNfluct}) reduces to 
\begin{equation}
  C_l^{\rm SN}={n_0L_0^2\over (4\pi)^2}
  \int^{\etac}_{\eta_i}d\eta \Bigl(a^{1-e+\alpha}\Bigr)^2 a^{-d}
  {-K\over \sinh^2\Delta\eta}.
\label{Clshotn}
\end{equation}
Assuming $\etac\simeq\eta_0$ and 
$(\eta_0-\etac)/\eta_0\ll1$,
we approximately rewrite it as
\begin{eqnarray}
  &&C_l^{\rm SN}\simeq{n_0L_0^2\over (4\pi)^2}
  \int^{\etac}_{\eta_i}{d\eta \over (\eta_0-\eta)^2}
  \simeq{n_0L_0^2\over (4\pi)^2}{1 \over \eta_0-\etac},
\\
  &&\Sc\simeq {L_0 \over 4\pi(\eta_0-\eta_c)^2},
\label{Scsimeq}
\end{eqnarray}
where the second equation is derived from (\ref{Scut}). 
Then we have 
\begin{equation}
  C_l^{\rm SN}\simeq \Sc^{1/2} n_0 \biggl({L_0\over 4\pi}\biggr)^{3/2}. 
\label{ClSN}
\end{equation}
This approximation is valid for the case that 
the removed bright sources are located at low redshift $z\ll 1$.
Using the value $I^{(0)}=5.2\times 10^{-8}{\rm erg~s^{-1} cm^{-2} str^{-1}}$
in $2-10~{\rm keV}$ band \cite{Boldt}, we have
\begin{equation}
  {\sqrt{C_l^{\rm SN}}\over I^{(0)}}
  =1.2 \biggl({\Sc\over {\rm erg~s^{-1} cm^{-2} }}\biggr)^{1/4}
  \biggl({L_0\over 3\times 10^{43} {\rm erg~s^{-1}}}\biggr)^{3/4}
  \biggl({n_0\over 3\times 10^{-5} {\rm Mpc^{-3}}}\biggr)^{1/2} .
\label{CSN}
\end{equation}
   From (\ref{Scsimeq}) and $\eta_0-\eta\simeq z/H_0$,
which is obtained from the Friedmann equation, we 
estimate the redshift $\zmin$ to which the bright sources 
are removed as
\begin{eqnarray}
  \zmin\simeq\biggl({L_0\over \Sc}{H_0^2\over 4\pi}\biggr)^{1/2}
  \simeq0.03h 
  \biggl({L_0\over 3\times 10^{43} {\rm erg~s^{-1}}}\biggr)^{1/2}
  \biggl({\Sc\over 3\times 10^{-11}{\rm erg~s^{-1} cm^{-2} 
  }}\biggr)^{-1/2}.
\end{eqnarray}

As we see from (\ref{ClSN}), the shot noise fluctuations are controlled
by the flux cutoff limit $\Sc$. When $\Sc$ is decreased the shot noise
fluctuations reduce too.  As we saw in the previous subsection, the
amplitude of the low multipole fluctuations due to the source clustering
decreases when $\Sc$ is decreased. However, the higher multipoles are
constant even when $\Sc$ is decreased.  The epoch-dependence of the bias
factor makes the amplitude of small-scale fluctuations significant.
Now let us compare these two fluctuations, the source clustering and the shot
noise.  In Fig.3 the fluctuations due to source clustering are compared
with the shot noise fluctuations as a function of $\Sc$. In this figure
we focused on the multipole of $l=100$.  The upper panel is the case of
no bias by setting $b_{X0}=1$ in equation (\ref{bX}). In the lower panel
we chose $b_{X0}=1.6$.  The large difference between these two panels,
which is due to the high bias factor at high redshift, implies that the
bias is a very important factor besides the evolution of the luminosity
function of sources when predicting the fluctuations due to source clustering.

\section{Summary and Discussion}
In this paper we have formulated equations to investigate the X-ray
background fluctuations in various cosmological models in a systematic
way.  We have made full use of the harmonic expansion method in terms of 
multipole components in both open (hyperbolic) and flat universes to
obtain the angular power spectrum of fluctuations.  The fluctuations due
to the source clustering and the shot noise can be predicted if the
evolution of the luminosity function and the bias factor are given.  We
have calculated the angular power spectrum for the X-ray background
fluctuations in various cosmological models based on a simple X-ray
source model.  As Lahav, et al. pointed out \cite{Lahav}, the
fluctuations predicted in the simple model strongly depend on the
evolution parameter of sources, and rather weakly depend on the
cosmological models.  Moreover the large-angle (low multipole of $l\lsim10$)
fluctuations are strongly  affected by the flux cutoff parameter for removal of
nearby bright sources~\cite{Miura,Lahav}. This is because the nearby
sources of low redshift are the dominant contributors to the large-angle
fluctuations.  Therefore when one makes the comparison between the
theoretical prediction and the observational data on large angular
scales, the procedure of removing bright sources has to be very
carefully taken into account.


In this paper, we have developed the formulation which is 
useful to calculate the angular power spectrum. 
In the observational point of view, however, 
we may not have to derive $C_\ell$ in small scales\cite{Barcons,Treyer}. 
The description by $\big<C(\theta)\big>$ is usually used
when one compares the theoretical prediction with
observational data (e.g., \cite{DeZotti,TL}).
Of course, we can easily derive $\big<C(\theta)\big>$ from $C_\ell$
by equation (\ref{Cl}) with a window function.
Moreover, our approach is useful to understand 
the physical mechanism which determines the behaviors of 
fluctuations on various scales.
This formulation is not limited to the X-ray background
fluctuations but it can be easily generalized to 
calculate the clustering of point-like sources
at high redshift.

In predicting the X-ray background fluctuations the evolution
of luminosity function is an important factor.
Our analysis shows that the evolution of the bias factor
is also a very important factor especially on small scales.
This is because the small-scale fluctuations contain information 
of clustering in the high redshift universe.
In order to use the X-ray background fluctuations as a probe of 
large scale cosmological density fluctuations at high redshift, 
the knowledge of the evolution of the bias factor and the luminosity 
function is necessary factor, because the  
dependence of cosmology
on the X-ray background fluctuations is rather weak.
Conversely, which suggests that the X-ray background
is a good probe for the bias mechanism of high redshift sources
\cite{JF,Tegmark} when the cosmological parameters are fixed.
The future X-ray mission projects will give us fine
solution for the evolution of X-ray sources.
In that case the X-ray background will be a
possible probe to investigate the clustering and
formation process of X-ray sources in the high redshift universe. 

\begin{center}
{\bf ACKNOWLEDGMENT}
\end{center}
One of the authors (K.Y.) thanks Prof. Y.Kojima for comments 
and encouragement. 
He is grateful to Prof. K.Tomita and J.Yokoyama and people in Yukawa 
Institute for Theoretical Physics, Kyoto university, where part of the 
work was done. He also thanks Prof. Y.Suto, N.Gouda and O.Lahav for 
useful comments and discussions.
We acknowledge helpful conversations on the subject of this paper
with Prof. M.Sasaki, K.Mitsuda, T.Matsubara, T.Ohsugi, N.Sugiura 
and Y.Miura.
This work is supported by the Grants-in-Aid for Scientific Research
of Ministry of Education, Science and Culture of Japan (No.09740203). 

\newpage
\begin{center}
{\bf FIGURE CAPTION}
\end{center}
\noindent
{Fig.~1--- The angular power spectrum of fluctuations due to source
  clustering for various cosmological models.  The cosmological
  parameters are taken as $h=0.5$ and $\Omega_0=1$ for the SCDM model, and
  $h=0.7$ and $\Omega_0=0.3$ for the $\Lambda$CDM and OCDM models.  The
  parameters $p\equiv d+e=3$ and $\zmax=3$ are taken.  In each panels
  three lines correspond to $z_{\rm min}=0,~0.02,~0.1$, respectively.
  Fig.1(a) is the case of no bias by setting $b_{X0}=1$ in equation
  (\ref{bX}). In Fig.1(b) $b_{X0}=1.6$ is chosen as a case of
  epoch-dependent bias.  Note that the area under the lines do not
  directly describe the amount of observed fluctuations since
  $\sqrt{\ell(2\ell+1)C_\ell}$ is the one in case of the logarithmic
  interval of x-axis (see Eq.~\ref{Cl}).  Roughly speaking, we need to
  multiply $\ell$.  Therefore there are larger fluctuations on smaller
  scales. On the other hand, the contribution from the shot noise does
  not depend on $\ell$ in these figures.}

\vspace{0.5cm}
\noindent
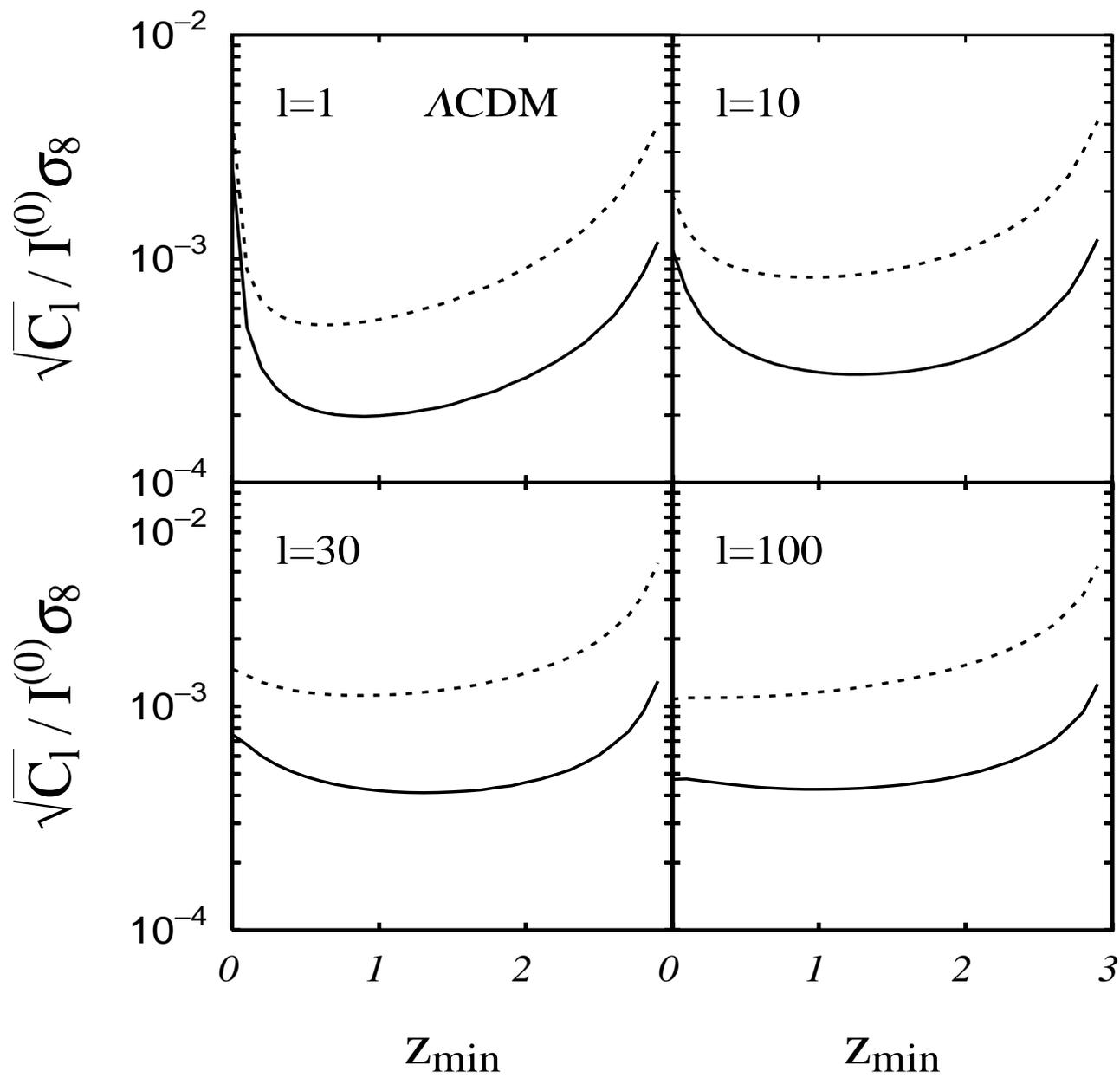
{Fig.~2--- $z_{\rm min}$ dependence of multipoles
for the $\Lambda$CDM model. The four panels correspond 
to the multipoles $l=1,~10,~30,~100$, respectively. 
The solid line represents the case $b_{X0}=1$ and the dashed 
line does the case $b_{X0}=1.6$. The model parameters
are same as that in Fig.1(a).

\vspace{0.5cm}
\noindent
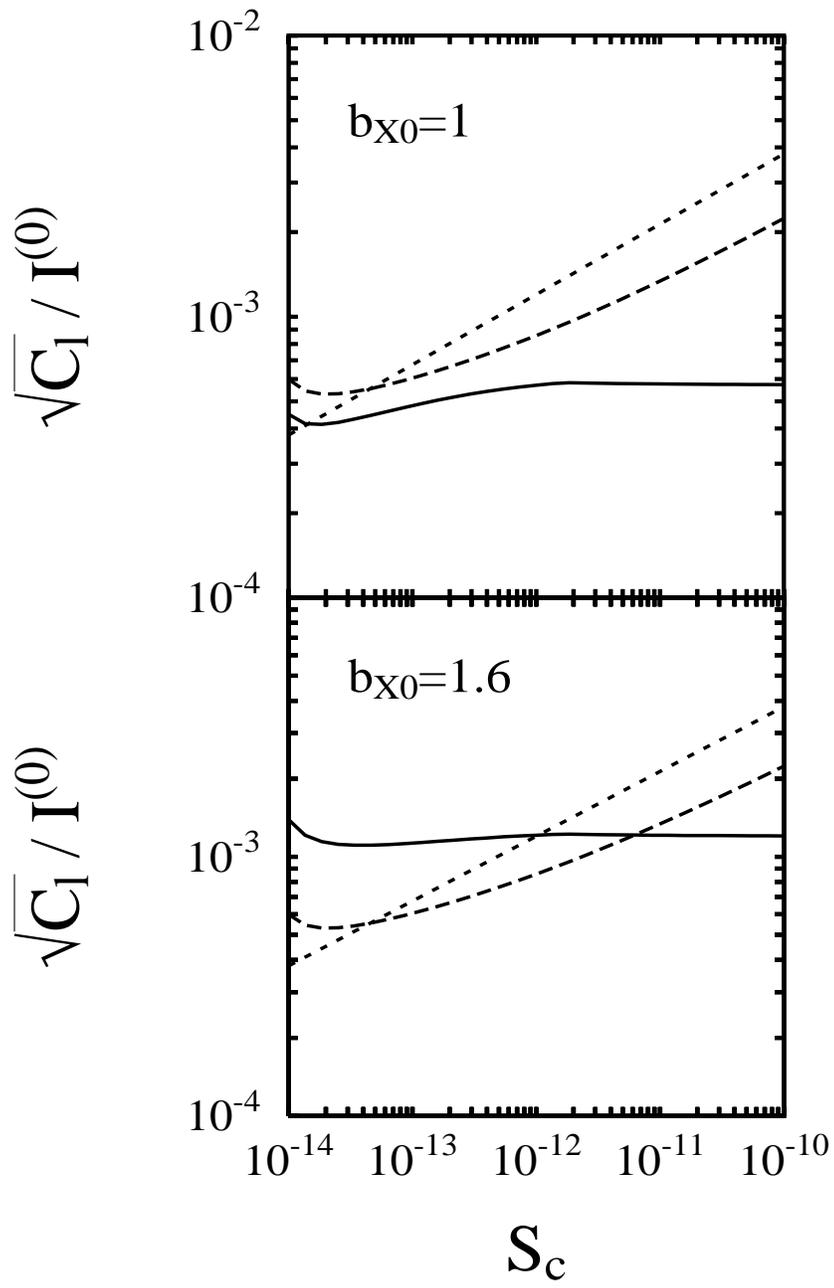
{Fig.~3--- Comparison of fluctuations
due to source clustering and shot noise term
as a function of flux cutoff limit $\Sc$.
The upper panel is the case of no bias by setting 
$b_{X0}=1$ in equation (\ref{bX}). In the lower panel  
$b_{X0}=1.6$  is chosen as a case of epoch-dependent bias.
Each panel shows the multipole $l=100$.
The solid line shows the clustering fluctuations
and the dashed line does the shot noise fluctuations.
For the clustering fluctuations we used the SCDM model
with $h=0.5$ and $\Omega_0=1$.
Here we normalized the CDM density fluctuations by $\sigma_8=0.5$.
In order to calculate the shot noise term, 
we need to specify the values of $e$ and $d$, separately.
Though there are many uncertainties in the evolution model of the 
sources, let us take the values $e=1.4$ and $d=1.6$.
In this case we have $1+\alpha-e=0$, which allow us
some analytic calculations when solving equation (\ref{Scut}).
We also used the parameters $L_0=3\times10^{43} {\rm erg/s}$,
$n_0=10^{-5}/{\rm Mpc^3}$. We set $\zmax=3.2$ in order to
be $I^{(0)}=5.2\times 10^{-8}{\rm erg/s/cm^2}$ \cite{Boldt}.
In this case of the parameters the estimation for the shot noise 
fluctuation is $\sqrt{C_l^{\rm SN}}/I^{(0)}\simeq 0.7 \Sc^{1/4}/({\rm
erg/s/cm^2})$.
The dashed line shows the result
obtained by calculating Eq.(\ref{Clshotn}).
The dotted line is the result obtained by extrapolating
the estimation by Lahav et al.\cite{Lahav}.}
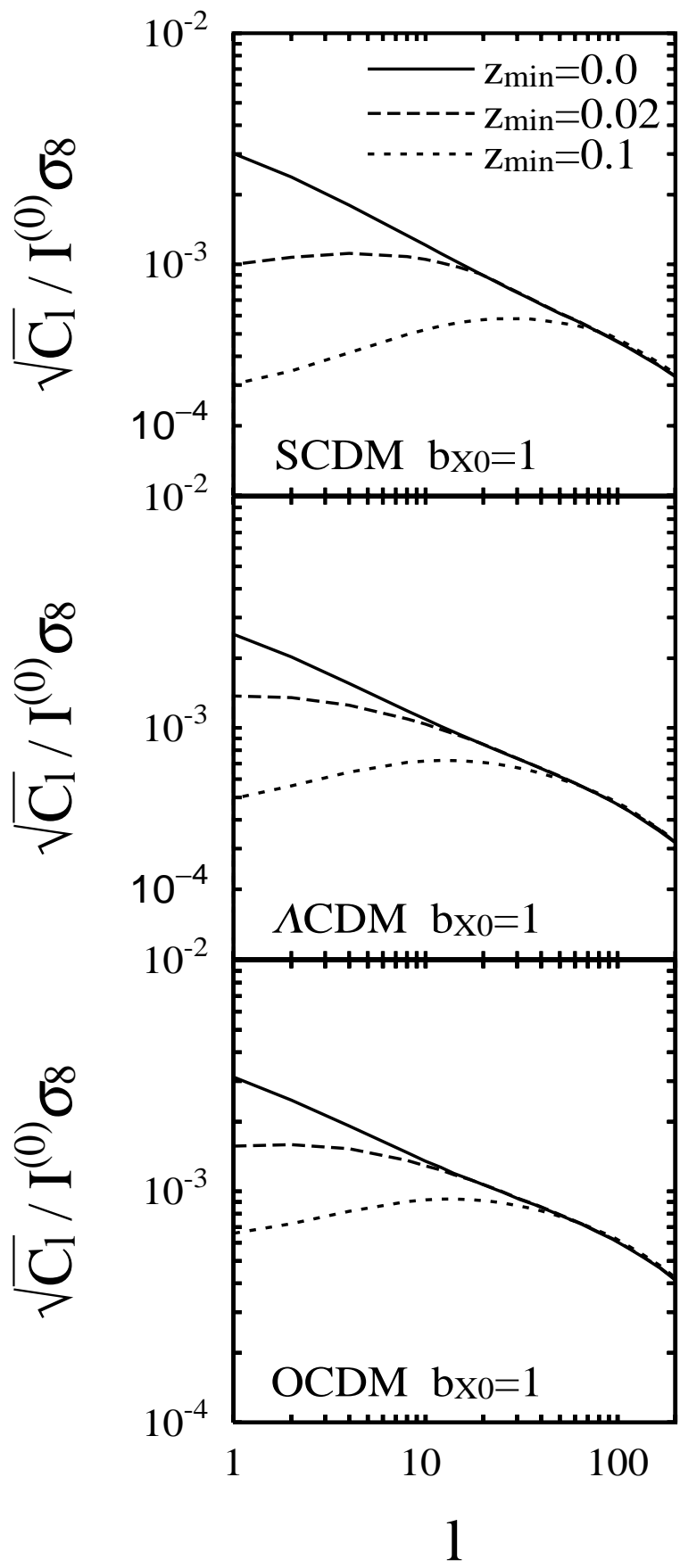
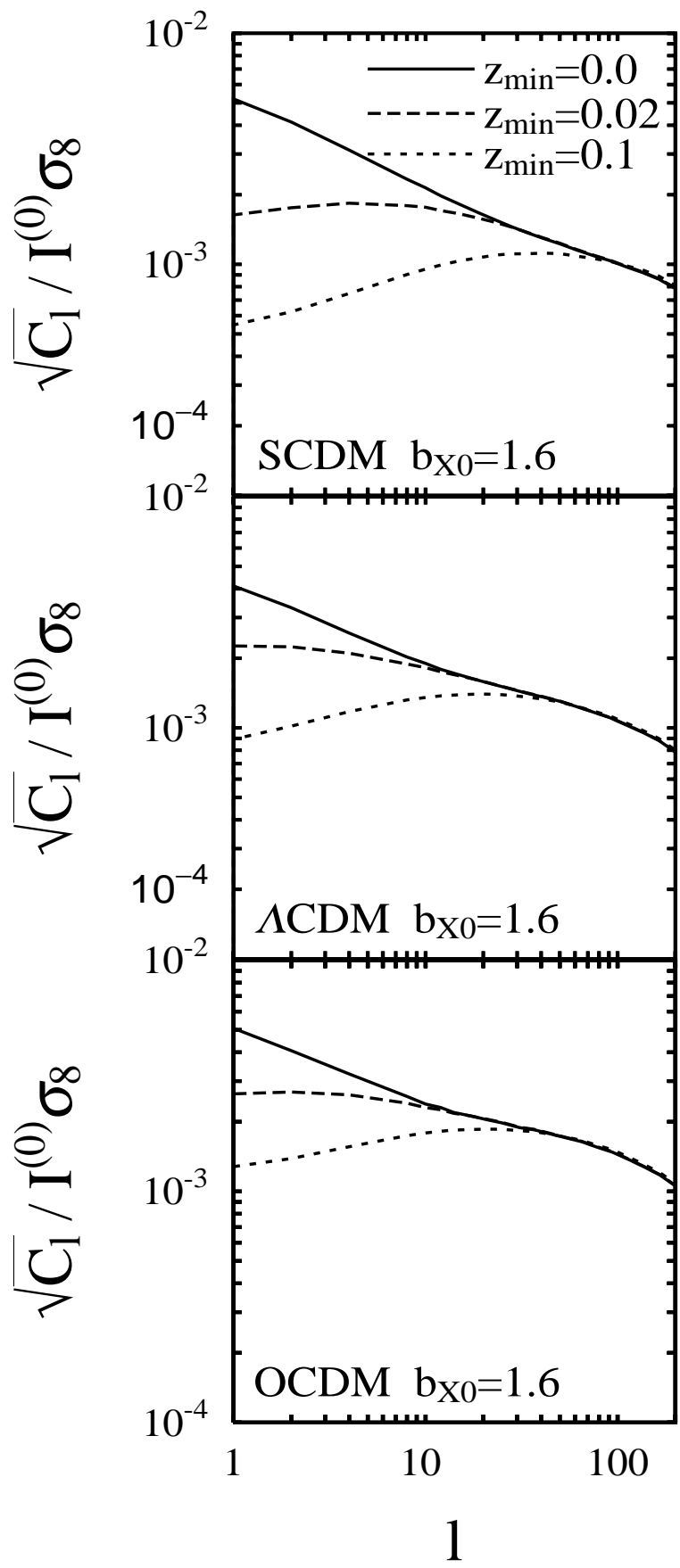
\end{document}